# SpecTel: A 10-12 meter class Spectroscopic Survey Telescope

Astro2020 Decadal Survey facilities white paper

Project: Large Ground Based (>$70M); Multi-Object Spectroscopic Facility

Richard Ellis (UC London, European Lead): richard.ellis@ucl.ac.uk
Kyle Dawson (University of Utah, U.S. Lead): kdawson@astro.utah.edu

Joss Bland-Hawthorn (Sydney), Roland Bacon (Lyon), Adam Bolton (NOAO), Malcolm Bremer (Bristol), Jarle Brinchmann (Porto), Kevin Bundy (University of California Observatory), Charlie Conroy (Harvard CfA), Bernard Delabre (ESO), Arjun Dey (NOAO), Alex Drlica-Wagner (FermiLab, University of Chicago), Jenny Greene (Princeton University), Luigi Guzzo (Milan), Jennifer Johnson (Ohio State University), Alexie Leauthaud (UC Santa Cruz), Khee-Gan Lee (Kavli IPMU), Luca Pasquini (ESO), Laura Pentericci (Rome), Johan Richard (Lyon), Hans-Walter Rix (Heidelberg), Connie Rockosi (UC Santa Cruz), David Schlegel (Lawrence Berkeley National Laboratory), Anže Slosar (Brookhaven National Laboratory), Michael Strauss (Princeton University), Masahiro Takada (Kavli IPMU), Eline Tolstoy (Groningen), Darach Watson (Copenhagen)

We recommend a conceptual design study for a spectroscopic facility in the southern hemisphere comprising a large diameter telescope, fiber system, and spectrographs collectively optimized for massively-multiplexed spectroscopy. As a baseline, we propose an 11.4-meter aperture, optical spectroscopic survey telescope with a five square degree field of view. Using current technologies, the facility could be equipped with 15,000 robotically-controlled fibers feeding spectrographs over 360<$\lambda$<1330 nm with options for fiber-fed spectrographs at high resolution and a panoramic IFU at a separate focus. This would enable transformational progress via its ability to access a larger fraction of objects from Gaia, LSST, Euclid, and WFIRST than any currently funded or planned spectroscopic facility. An ESO-sponsored study (arXiv:1701.01976) discussed the scientific potential in ambitious new spectroscopic surveys in Galactic astronomy, extragalactic astronomy, and cosmology. The US community should establish links with European and other international communities to plan for such a powerful facility and maximize the potential of large aperture multi-object spectroscopy given the considerable investment in deep imaging surveys.

## Key Science Goals and Objectives

Spectroscopy will always be a primary tool of astronomy. It yields unique insight into our local neighborhood by providing the chemical composition and radial velocities for stars in the Milky Way and nearby resolved galaxies. On larger scales, spectroscopy provides accurate redshifts for cosmology, measures of internal motions, the nature of stellar populations and non-thermal sources, and a census of the gaseous and stellar composition and ionizing radiation field over cosmic time. The Sloan Digital Sky Survey (SDSS) and surveys with the UK-Australian 2 degree field (2dF) instrument have clearly shown the benefit of combining imaging surveys with multi-object spectroscopic data matched in quality and quantity. These facilities were productive for several decades and modest reconfigurations with updated detectors and/or new instrumentation addressed new and previously unforeseen scientific questions. Of particular significance is the impact of panoramic integral field unit spectrographs such as ESO's MUSE and Keck's KCWI, enabling survey spectroscopy without pre-assigned targets.

Deep imaging from the Large Synoptic Survey Telescope (LSST), Euclid, and WFIRST will provide new catalogues of spectroscopic targets in greater numbers and to much fainter limits than can be observed with any current spectroscopic facility. These imaging programs pave the way for a dedicated spectroscopic facility in the southern hemisphere that optimizes multiplexing power, collecting area, and field of view. A 10-12m class telescope can exploit these rich photometric datasets through optical and near infrared spectroscopy. By providing key diagnostic and local environmental data for targets observed at other wavelengths, and offering a multi-dimensional view of hundreds of millions of stars and galaxies, such a facility will enable transformational progress in Galactic and extragalactic astronomy.  Such a facility will significantly increase the legacy value of LSST. The scientific opportunities provided by such a facility align with numerous submitted Astro2020 white papers:

**Assembly history of the Milky Way and the role of dark matter:** Our Milky Way and its immediate environment offer a unique opportunity to understand how galaxies of a range of masses assembled and the role of dark matter in that process. Stars in the Milky Way and its satellites encode decisive information about galaxy formation and environmental processes. Spectroscopy of large samples of stars yielding ages, masses and orbits can probe the gravitational potential of the Milky Way; spectroscopic studies of stellar streams can trace the presence of starless dark matter halos from which a census can test the cold dark matter model and alternatives[1]. Through `chemical tagging', detailed stellar abundances act as a `fingerprint' identifying widely-dispersed stars of common origin and hence provide a route to reconstructing the assembly history of the Milky Way. Some nearby satellite galaxies formed their stars at very high redshift providing detailed insight into the physics of the early Universe.

To address these important questions requires precision velocities, proper motions (from Gaia) and abundances for a large range of elements, especially r-process elements[2,3,4], where high resolution spectroscopy at blue and UV wavelengths is required. The proposed facility is designed to dramatically improve the statistics in all the above areas through its increased aperture and multiplex gain.

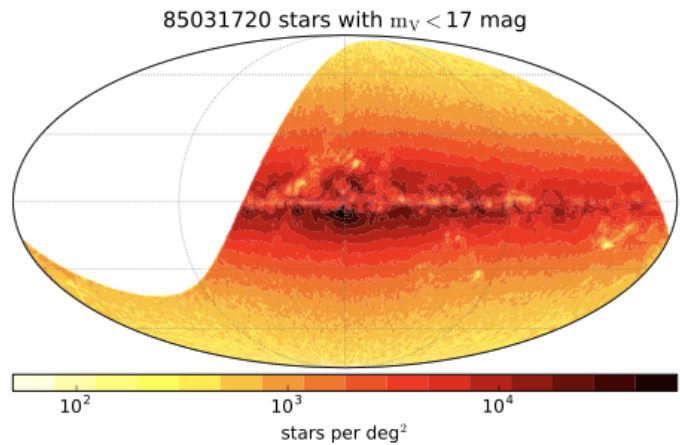

***Figure 1***: *Distribution of the 85 million stars brighter than V < 17 across the Southern sky, based on the Galaxia model; courtesy J. Rybitzki.*

An example survey would secure **high resolution** (R ~ 20,000 – 40,000) spectra at a S/N~80 for 30-85M stars to a magnitude limit of V~15.5 – 17 (Figure 1), providing a rich array of chemical and kinematic data across the southern Milky Way. Chemical tagging studies scale as the square of the sample size and linearly with the number of elements; the number of associated stars would increase 300-fold over that achievable with any currently-funded facility. With a 10-12m class telescope, exposure times of 1-2 hours would be sufficient, but a field of view of ~5 deg$^2$ is required to match the stellar density and exploit a large multiplex gain. For individual stars in nearby dwarf galaxies, the requirement is to measure accurate abundances to V>17.5. As ultra-faint dwarfs contain few red giant branch (RGB) stars, it is necessary to push toward the main sequence turnoff. Whereas dwarf spheroidals have populated RGBs, they are more distant (> 80 kpc). In both cases, the target areas exceed the field of view of current telescopes. Longer integrations for these dwarfs and stellar streams could be included as part of an all-sky survey. Securing R~40,000 spectra with S/N~30 at V ~19-20 would take 5 hours.

High signal-to-noise spectroscopy at **moderate resolution** (R~4000) can also yield accurate radial velocities and metallicities. Gaia is providing radial velocities for only 10% of its sources, leaving more than a billion stars with excellent astrometry and photometry but no spectroscopy. Spectroscopy of a large fraction of these would provide 3D positions and kinematics, stellar type, and metallicities. Characterization of stellar streams and the profile of the dark matter halo requires line-of-sight velocities to a precision of a few km/s[5]. Five years operation of the 5000-fiber DESI spectrograph may achieve these measures for 100 million g<20 stars[6]. A 15,000 fiber instrument on a 10-12-meter telescope could obtain such measurements for 650 million stars in only a year. Studies at fainter magnitudes would also be possible. By sampling RGB, blue horizontal branch, and RR Lyrae stars in the halo to r<22.5, a census of satellite galaxies, globular clusters, high velocity stars, and stellar streams can be achieved to over 100 kpc[7].

Galactic astronomy is undergoing a phenomenal renaissance, largely as a result of the Gaia mission. The proposed facility can exploit this by delivering kinematic and chemical abundance data over the entirety of the Milky Way and its associated satellites to place `near-field cosmology' on par with its more distant partner.

**Galaxy evolution in the context of the cosmic web and intergalactic medium:** Deep imaging surveys will, by 2025-2030, provide 0.2 arcsec image quality over more than 15,000 deg$^2$ from Euclid and multi-band photometry to AB~27 from LSST. Most spectra exploiting these data (e.g. from Euclid, DESI and PFS) will have low resolution and modest S/N. Higher quality spectra from JWST and ELTs will sample much smaller fields. The proposed facility will resolve these deficiencies with good spectral resolution and improved S/N. A good example of a panoramic spectroscopic strategy is SDSS which provided quality spectra for the z < 0.2 Universe. Although the initial motivation was cosmology and large scale structure, SDSS provided an enormous stimulus to the physics of the galaxy population. At earlier cosmic times extensive multi-band photometry is needed to ensure careful pre-selection in the targeted redshift range; this hurdle will be overcome with Euclid and LSST. Moreover, wide-field coverage with completeness exceeding 90% is essential to span the full range of environmental densities that influence galaxy assembly. Coverage from low redshifts to z>1 is required to characterize evolution[8]. A dedicated wide-field facility can open a new chapter on galaxy evolution studies.

The most fundamental feature of structure formation is the complex distribution of matter referred to as the *cosmic web* (Figure 2a). Galaxies grow in this evolving network of dark matter halos, filaments and voids, accreting and ejecting baryons. How do these structures on various physical scales affect accretion, star formation, and outflows in galaxies? What is the cycle of chemical enrichment in the intergalactic medium – are there regions that are chemically pristine? Mapping correlations between these structures and galaxy properties over 1<z<4 and scales that span the full range of web environments implies surveying ~Gpc$^3$ volumes in various redshift bins with dense sampling. Multi-object spectroscopy on such scales would enable detailed estimates of star-formation rates, AGN contributions, and dynamical mass estimates for galaxy clusters at a critical time of transition[9]. This formidable challenge can only be achieved with a dedicated wide-field 10-12m class facility.

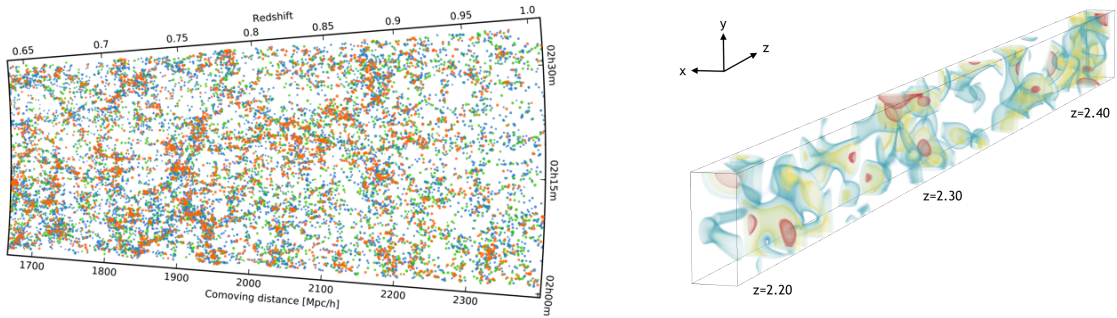

*Figure 2*: Charting the cosmic web: (Left:) A densely-sampled redshift survey such as in the VLT VIPERS survey where colors code quiescent and star forming galaxies over 0.6<z<1. The proposed facility would extend such surveys to z~2. (Right:) Reconstruction of the 3-D Lyman alpha absorption field observed in the spectra of multiple background Lyman break galaxies; colors trace environmental overdensities.

Reconstructing the evolving 3-D density distribution at z>2 has been demonstrated using Lyman-alpha forest absorption spectra of background Lyman break galaxies (LBGs) along multiple lines of sight (https://arxiv.org/abs/1409.5632, Figure 2b). Overdensities so derived also provide a map of protoclusters that can be correlated with galaxy maps to measure the relationship between underlying dark and baryonic structure[10]. Current Lyman-alpha tomographic surveys over ~0.2 deg$^2$ achieve a transverse sampling of ~2.3$h^{-1}$ Mpc (https://arxiv.org/abs/1710.02894). The logical next step is to exploit this technique over representative cosmic volumes. The fluxes and surface densities required to chart evolution using LBGs as tracers and the Ly-alpha forest as a probe of dark matter are similar. To span ~1 Gpc$^3$ implies covering ~200 deg$^2$ areas from AB~23 at z~1 to AB~25 at z~4. Current 10m spectroscopic surveys at these depths indicate exposure times from 2 hours to 5-7 hours respectively. Charting six Gpc$^3$ volumes over 1 < z < 4 would take 400 dark nights.

A large aperture facility offers higher spectral resolution (e.g. R~3000) to separate absorption lines of the interstellar gas from stars and provide key measures of outflow kinematics and chemical composition through UV abundance measurements[11]. A large survey can resolve Lya emission line profiles as probes of the escape of ionizing radiation. This more ambitious absorption line survey can be conducted on a subset of AB<24 galaxies with exposure times of ~20-50 hours. Targeting ~5% of the galaxies in the earlier cosmic web survey would lead to 2 x 10$^5$ galaxies with exquisite spectra – leading to a ``SDSS@z~2-3'' two orders of magnitude larger than the VANDELS survey (http://vandels.inaf.it/. The above surveys illustrate the astonishing power of combining a 10-12 meter aperture with a panoramic field and high multiplex gain. Both surveys could readily be concluded over a 3-5 year timespan – an important criterion is defining what is practical. Such surveys could not be accomplished with the upcoming generation of renewed 4m facilities (DESI, 4MOST, WEAVE) and neither MOONS nor PFS offers a dedicated facility. Alongside traditional multi-fiber spectroscopic surveys based on targets pre-selected from deep photometry, a panoramic IFU offers the opportunity for statistically-complete deep emission line and other serendipitous searches as evidenced by the significant impact of ESO's MUSE instrument.

**Cosmology:** Dark energy represents the most important mystery in both particle physics and cosmology and will remain a focal point even after the completion of DESI, PFS and LSST. Improving statistical and systematic precision in dark energy constraints by taking more data, developing more advanced probes, and considering modified gravity and the full suite of unanswered questions in fundamental physics will likely motivate a vibrant cosmology program well into the 2030s[12]. LSST will image the southern, extragalactic sky in six filters providing a photometric catalog of > 4 billion galaxies. Future advances will rely upon precision redshifts for a large, carefully-selected sample of these galaxies. There are at least three regimes of cosmological research that require a massively-multiplexed spectroscopic facility. Each of these three expands and complements the galaxy evolution science cases:

(1) Photometric redshifts are critical to LSST studies with weak lensing and other techniques. For photometric redshift errors to be a subdominant source of uncertainty, a sample of at least 20,000 galaxies to I~25.3 spanning the full color and magnitude range is required[13]. A spectroscopic sample obtained over 15 independent fields with S/N similar to that of DEEP2 would take roughly seven months of dark time. (http://d-scholarship.pitt.edu/36036/7/Photo-z%20training%20survey%20times.pdf).
Observations of the faintest sources could be coordinated with the 20-50 hour absorption line survey of AB<24 galaxies discussed above.

(2) A densely-sampled redshift survey can explore higher-order statistics in clustering[14], obtain precise estimates of structure growth via cross-correlations with weak lensing maps[15], identify a complete sample of voids[16], and extract new cosmological information from non-linear clustering. A sample of 20,000 galaxies deg$^{-2}$ could be color-selected from LSST images to precisely-tuned comoving number densities to redshifts z<1.5, yielding clustering information to scales a factor of two smaller than those typically considered to define the linear regime. Observing such a sample over 14,000 deg$^2$ with one hour integrations would lead to a program with high redshift completeness and would require about 280 million fiber-hours of dark time.

(3) The cases for dark energy[17], inflation[18], neutrino[19], complex models of dark matter[20], and primordial non-Gaussianity[21] studies were presented in many Astro2020 white papers. These objectives are best pursued over the large volumes at high redshift (1.5<z<4) using 3-D clustering of at least one hundred million galaxies, a sample significantly larger than DESI's at these redshifts. LSST imaging will provide a sample of candidate Lyman-break and Lyman-alpha emitting galaxies at z>2 for these studies. A sample that targets 15,000 galaxies deg$^{-2}$ at g<24.5 over a 14,000 deg$^2$ would contain most of the cosmological information from the linear regime. Assuming one hour integrations to obtain precise redshifts of the Lyman-alpha emitting sub-population and those galaxies that are sufficiently bright to detect multiple absorption lines would require about 200 million fiber-hours of dark time.

**The Transient Universe:** LSST is expected to find more than a million transient events in its first five years, leading to an exciting growth area in astrophysics. Spectroscopic follow up will be key to characterize these sources and uncover new discoveries. The challenge is illustrated by considering supernovae (SNe). LSST will likely discover 300,000 SNe per year (65% Type Ia, 35% Type II). Assuming these are visible for ~50 days, the surface density of "live" transients is still only 2 deg$^{-2}$ at any instant. For the proposed panoramic facility, only ten live events per field could be followed up in real time. However, since a galaxy survey can readily study ten fields a night to AB~23, roughly 10% of all visible LSST transients could be observed via merging fiber allocations "on the fly". This would lead to spectra of more than 20,000 SNe Ia for every year of spectroscopy that overlaps LSST observations – a major step forward in precision studies of the Hubble diagram and using SNe Ia as a probe of weak lensing magnification[22]. The follow-up of "transpired" events for which LSST previously delivered exquisite photometric light curves and classifications will characterize much larger samples. Targeting the host galaxies for redshifts and other properties may offer the best route for studies of lower redshift SNe Ia such as those that probe structure growth through peculiar velocities [23].

The discovery space of LSST will surely lie in unexpected or rare transients. Although these do not drive the design of the proposed facility, it will be advantageous to include them in survey target allocations in real time. Important targets include stripped stars, superluminous SNe, gravitational wave counterparts and kilonovae. Together with reverberation mapping of thousands of AGN[24], a dedicated spectroscopic facility can become the default follow-up machine for the transient community by offering "target of opportunity" modes within pre-determined galaxy surveys.

## Technical Overview

A well-planned and flexible survey facility can provide cutting edge science for decades if conceived as an end-to-end project.  This is essential given the instrumental cost is likely comparable to the telescope and dome.  We describe the technical components of a baseline spectroscopic facility designed in response to the ESO study.  The architecture is optimized for wide field, multi-object spectroscopy with the capability of a panoramic IFU at a separate focus and flexibility for future upgrades[25,26].

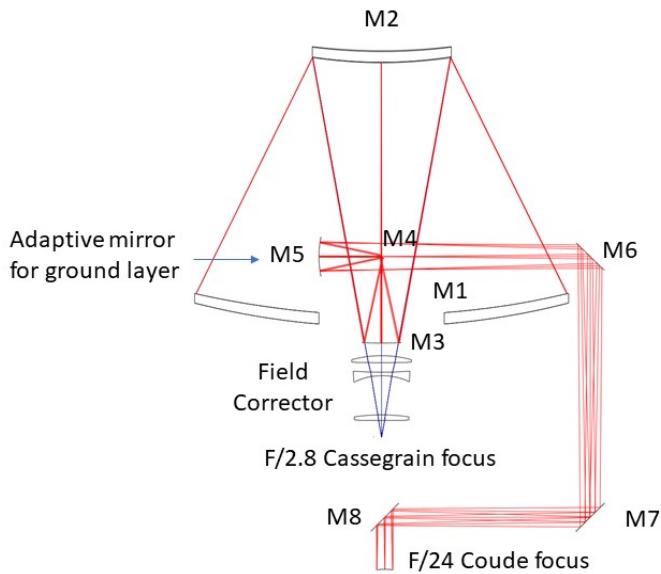

*Figure 3: Telescope with Cassegrain fiber focus and six arcmin FOV Coude focus for an IFU. Image quality was computed at zenith distances of 0, 30, and 55 deg. Cassegrain images are polychromatic between 360 and 1200 nm with RMS diameter ranging from 0.3 to 0.7 arcsec. The ADC lenses are made from Fused Silica, leading to a mild mismatch in refractive indices and reduced performance at $\lambda$ >1200 nm.The coude focus images are better than 0.1 arcsec; the 800 mm concave mirror at the elevation axis can be adaptive to correct the ground layer.*

**Telescope:** The 11.4 meter diameter telescope is designed to access the largest possible focal plane while maintaining a plate scale and input beam well-suited for fibers (Figure 3). The primary mirror consists of 78 ELT primary segments. The secondary mirror is 4.2m in diameter, similar to the ELT secondary. The three-lens corrector (1.8 m diameter largest lens) works also as Atmospheric Dispersion Compensator (ADC) and provides a corrected 2.5 degrees, 1.43m diameter Cassegrain focal plane at F/2.86 and a platescale of 570 mm/deg.

| Facility | Telescope Diameter (m) | Surface Area (m$^2$) | Field of view (deg$^2$) | Multiplex Number |
|---|---|---|---|---|
| MAYALL DESI | 3.8 | 9.6 | 8 | 5,000 |
| SUBARU PFS | 8.0 | 48.75 | 1.33 | 2,400 |
| VLT MOONS | 8.0 | 48.75 | 0.136 | 1,000 |
| MSE | 11.2 | 96.0 | 1.5 | 4,000 |
| **SpecTel** | **11.4** | **87.89** | **4.91** | **15,000** |

The Cassegrain configuration allows the maximum field of view and focal plane area: primary requirements for a system optimized for multi-object spectroscopy. The overall design is compact, minimizing the dimensions of the enclosure. The focal plane is easily accessible, allowing a fiber system that preserves transmission through shorter fibers. The focal plane requires only a three-lens corrector+ADC, thus reducing the number of coating surfaces, scattered light, and other effects that cause light loss. The combined FOV and effective area results in a facility more powerful than any currently funded, as shown in Table 1. By inserting two mirrors ahead of the corrector, the central 10 arcminutes can be directed to a gravity-invariant Coude focus hosting a panoramic IFU. The floor below the telescope rotates to compensate for the IFU field rotation.

**Focal Plane:** The 1.43m diameter Cassegrain focal plane is as large as possible given current limits in the manufacture of corrective optics. The focal plane allows a large number of fiber positioners while the plate scale allows relaxed constraints for centering c.f. DESI or 4MOST. As the focal plane has 3.1 times the area of DESI's prime focus, it would be possible to duplicate DESI's fiber positioner design to produce a 15,000 fiber system, although an improved design could be considered in the conceptual study.

**Fibers and Spectrographs:** The plate scale translates to 0.8 arcsec diameter for the 127 micron fibers employed on Subaru PFS. The PFS spectrograph designs could thus be employed with only minor modifications to account for the F/2.8 input beam. An alternative design would be one similar to that used in MOONS, e.g. Figure 4. Each camera consists of two aspherical lenses and an aspherical mirror, with an obscuration of roughly 20% caused by the detector housing. 600 fibers of 160 micron diameter (1 arcsec projected aperture) feed the spectrograph through a 300 mm diameter collimating mirror. The IR arm covers 965 to 1330 nm with R~4000, with some loss of throughput due to the degraded ADC performance at wavelengths longer than 1200 nm. There is sufficient room between the slithead and the VPH grating to insert dichroics feeding up to three optical arms. Two optical channels with R~3000 or three optical channels with R~4000 would extend coverage to 360 nm. Optical designs could follow a design similar to that presented for the IR channel or could follow a design with smaller pupil and less expensive optics if curved CCDs are implemented (see discussion in "Technology Drivers"). High resolution spectrographs would naturally be considered as part of the conceptual design process.

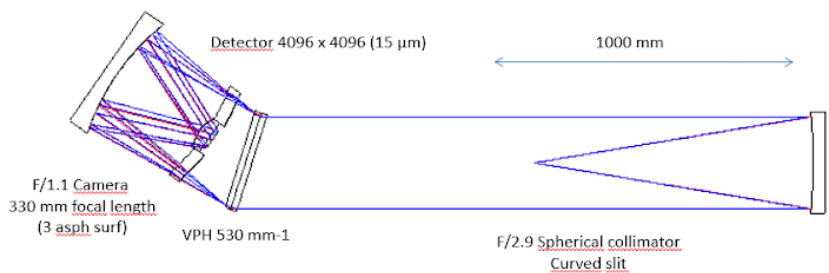

**Figure 4:** *Collimator, dispersive element, and achromatic, three element catadiotric IR camera. The camera provides R~4000 spectra from 965 to 1330 nm.*

**Technology Drivers**

A number of developments would improve performance, reduce cost, and increase the multiplex power. We encourage exploration of the following technologies that would lead to significant improvements over the baseline facility described above.

**Miniaturized electric motors for a higher density of fiber positioners:** Positioners at a pitch (packing) of 5 mm would allow a 4X increase in fiber number, thus enabling instrumentation of 60,000 fibers to pursue the full low- and high-redshift cosmology projects described in the "Key Science Goals and Objectives" section. Several fiber positioner designs show significant promise for smaller pitch systems[27].

**Germanium CCDs:** Infrared InGaAs and HgCdTe CMOS detectors have been used in ground- and space-based observatories, but they are expensive, require substantial cooling, and suffer from low fabrication yields. Germanium CCDs can be processed with the same tools used to build silicon devices, show promise for similar read noise and sensitivities and offer a high quantum efficiency to 1.4 microns[27].

**Curved CCDs or CMOS detectors:** A major cost of the spectrograph system is the large number of optical components with minimal aberration. This can be mitigated using detectors with curved surfaces matched to a fast camera. These detectors would likely be fabricated in a conventional foundry and molded to a curved mounting surface during the packaging process. Development of these detectors requires techniques to maintain a uniform electric field and pure substrate lattice after packaging.

### Organization, Partnerships, and Current Status

The baseline spectroscopic facility was largely motivated by the requirements of an ESO working group in response to community needs. The Australian Decadal Survey places such a spectroscopic facility at high priority and the Mauna Kea Spectroscopic Explorer (MSE) is an advanced concept facility proposed to replace the CFHT 4-meter telescope. The science cases described here span the missions of NSF and DOE. We encourage a model that leverages NOAO, Lab, and international expertise while seeking financial/in-kind contributions from NSF, DOE and international agencies.

### Schedule

The conceptual design study would establish a cost and construction schedule. We propose prioritizing the construction by relying on current technologies. A first stage might comprise the proposed telescope, 15,000 fiber system, and moderate-resolution spectrographs built in time for several years of coordinated observations with LSST. High-resolution spectrographs and a large IFU at the Coude focus would follow. This facility would be able to complete the Galactic program, the galaxy evolution and Lyman-alpha forest tomography programs, and the photometric redshift calibration program by the time that LSST completes. By the end of LSST, imaging with sufficient u-band depth will be available to select 15,000 high-redshift targets $\deg^{-2}$. Technological advances will likely increase the multiplex gain while decreasing the cost per spectrum. These advances could motivate a final stage with 60,000 fibers that is capable of completing a full sampling of the cosmic density field to high redshift (cosmology cases 2 and 3) while also building upon the stellar and galaxy evolution cases. This would remain the most powerful spectroscopic facility until at least 2050.

### Cost Estimates

A detailed cost estimate can only be obtained following a complete conceptual design study. Focusing on the major components of the facility, we find a total construction cost (FY19$) for the first-stage facility of ~$400M, based on an assumed cost of $200M for the telescope and enclosure with an equal amount for the spectrographs and fiber systems. An upgrade to a 60,000 fiber facility would cost an additional $600 million. R&D to include less expensive Germanium CCDs, easier assembly of fiber positioners, and/or curved detectors could substantially reduce costs in the upgraded system.

# Aligned Astro2020 White Papers and Related Papers

## Assembly history of the Milky Way and the role of dark matter

[1] Bechtol, et al., "Dark Matter Science in the Era of LSST",
  https://baas.aas.org/wp-content/uploads/2019/05/207_bechtol.pdf
[2] Roederer, et al., "The astrophysical r-process and the origin of the heaviest elements",
  https://baas.aas.org/wp-content/uploads/2019/05/136_roederer.pdf
[3] Rich, et al., "The Chemical/Dynamical Evolution of the Galactic Bulge",
  https://baas.aas.org/wp-content/uploads/2019/05/544_rich.pdf
[4] Johnson, et al., "The Origin of the Elements Across Cosmic Time",
  https://baas.aas.org/wp-content/uploads/2019/05/463_johnson.pdf
[5] Li, et al., "Dark Matter Physics with Wide Field Spectroscopic Surveys",
  https://baas.aas.org/wp-content/uploads/2019/05/252_li.pdf
[6] Dey, et al., "Mass Spectroscopy of the Milky Way",
  https://baas.aas.org/wp-content/uploads/2019/05/489_dey.pdf
[7] Sanderson, et al., "The Multidimensional Milky Way",
  https://baas.aas.org/wp-content/uploads/2019/05/347_sanderson.pdf

## Galaxy evolution in the context of the cosmic web and intergalactic medium

[8] Behroozi, et al., "Empirically Constraining Galaxy Evolution",
  https://baas.aas.org/wp-content/uploads/2019/05/125_behroozi.pdf
[9] Mantz, et al., "The Future Landscape of High-Redshift Galaxy Cluster Science",
  https://baas.aas.org/wp-content/uploads/2019/05/279_mantz.pdf
[10] Kartaltepe, et al., "Assembly of the Most Massive Clusters at Cosmic Noon",
  https://baas.aas.org/wp-content/uploads/2019/05/395_kartaltepe.pdf
[11] Smith, et al., "The Chemical Enrichment History of the Universe",
  https://baas.aas.org/wp-content/uploads/2019/05/400_smith.pdf

## Cosmology

[12] Slosar, et al., "Dark Energy and Modified Gravity",
  https://baas.aas.org/wp-content/uploads/2019/05/097_slosar.pdf
[13] Newman, et al., "Deep Multi-object Spectroscopy to Enhance Dark Energy Science from LSST",
  https://baas.aas.org/wp-content/uploads/2019/05/358_newman.pdf
[14] Wang, et al., "Illuminating the dark universe with a very high density galaxy redshift survey over a wide area",
  https://baas.aas.org/wp-content/uploads/2019/05/508_wang.pdf

[15] Mandelbaum, et al., "Wide-field Multi-object Spectroscopy to Enhance Dark Energy Science from LSST",
    https://baas.aas.org/wp-content/uploads/2019/05/363_mandelbaum.pdf

[16] Pisani, et al., "Cosmic voids: a novel probe to shed light on our Universe",
    https://baas.aas.org/wp-content/uploads/2019/05/040_pisani.pdf

[17] Ferraro, et al, "Inflation and Dark Energy from spectroscopy at $z > 2$",
    https://baas.aas.org/wp-content/uploads/2019/05/072_ferraro.pdf

[18] Slosar, et al., "Scratches from the Past: Inflationary Archaeology through Features in the Power Spectrum of Primordial Fluctuations",
    https://baas.aas.org/wp-content/uploads/2019/05/098_slosar.pdf

[19] Dvorkin, et al., "Neutrino Mass from Cosmology: Probing Physics Beyond the Standard Model",
    https://baas.aas.org/wp-content/uploads/2019/05/064_dvorkin.pdf

[20] Gluscevic et al., "Cosmological Probes of Dark Matter Interactions: The Next Decade",
    https://baas.aas.org/wp-content/uploads/2019/05/134_gluscevic.pdf

[21] Meerburg, et al., "Primordial Non-Gaussianity",
    https://baas.aas.org/wp-content/uploads/2019/05/107_meeburg.pdf

**The Transient Universe**

[22] Scolnic, et al., "The Next Generation of Cosmological Measurements with Type Ia SNe",
    https://arxiv.org/pdf/1903.05128.pdf

[23] Kim, et al., "Testing Gravity Using Type Ia Supernovae Discovered by Next-Generation Wide-Field Imaging Surveys",
    https://baas.aas.org/wp-content/uploads/2019/05/140_kim.pdf

[24] Shen, et al., "Mapping the Inner Structure of Quasars with Time-Domain Spectroscopy",
    https://baas.aas.org/wp-content/uploads/2019/05/274_shen.pdf

**Related Papers**

[25] Pasquini, Delabre, Ellis, and de Zeeuw, "New telescope designs suitable for massively-multiplexed spectroscopy",
    https://arxiv.org/abs/1606.06494

[26] Pasquini, et al., "The ESO Spectroscopic facility",
    https://arxiv.org/abs/1708.03561

[27] Dawson, et al., "Cosmic Visions Dark Energy: Small Projects Portfolio",
    https://arxiv.org/abs/1802.07216